\documentclass{jaa}
\usepackage{natbib}
\usepackage{ulem}
\usepackage{xcolor}
\DeclareRobustCommand{\VAN}[3]{#2}
\let\VANthebibliography\thebibliography
\def\thebibliography{\DeclareRobustCommand{\VAN}[3]{##3}\VANthebibliography}
\usepackage{subcaption}
\usepackage{graphicx}
\usepackage[flushleft]{threeparttable}
\begin{document}\sloppy

\title{\textit{AstroSat} observation of rapid Type-I thermonuclear burst from the low mass X-ray binary GX 3+1}

\author{ANKUR NATH\textsuperscript{1,2}, BIPLOB SARKAR\textsuperscript{1,*}, JAYASHREE ROY\textsuperscript{3} and RANJEEV MISRA\textsuperscript{3}}
\affilOne{\textsuperscript{1}Department of Applied Sciences, Tezpur University, Napaam-784028, Tezpur, Assam, India\\}
\affilTwo{\textsuperscript{2}Faculty of Science \& Technology, The ICFAI University Tripura, Agartala, 799210, India\\}
\affilThree{\textsuperscript{3}Inter-University Centre for Astronomy and Astrophysics (IUCAA), Post Bag 4, Ganeshkhind, Pune 411 007, India\\}

\twocolumn[{

\maketitle

\corres{biplobs@tezu.ernet.in}

\msinfo{XX XXX 20XX}{XX XXX 20XX}

\begin{abstract}
We report the results of an observation of low mass X-ray binary GX 3+1 with {\it AstroSat}'s Large Area X-ray Proportional Counter (LAXPC) and Soft X-ray Telescope (SXT) instruments on-board for the first time. We have detected one Type-1 thermonuclear burst ($\sim$ 15 s) present in the LAXPC 20 light curve, with a double peak feature at higher energies and our study of the hardness-intensity diagram reveals that the source was in a soft banana state. The pre-burst emission could be described well by a thermally Comptonised model component. The burst spectra is modelled adopting a time-resolved spectroscopic method using a single color blackbody model added to the pre-burst model, to monitor the parametric changes as the burst decays. Based on our time-resolved spectroscopy, we claim that the detected burst is a photospheric radius expansion (PRE) burst. During the PRE phase, the blackbody flux is found to be approximately constant at an averaged value $\sim$ 2.56 in $10^{-8}$ ergs s$^{-1}$ cm$^{-2}$ units. On the basis of literature survey, we infer that \textit{AstroSat}/LAXPC 20 has detected a burst from GX 3+1 after more than a decade which is also a PRE one. Utilising the burst parameters obtained, we provide a new estimation to the source distance, which is  $\sim$  9.3 $\pm$ 0.4 kpc, calculated for an isotropic burst emission. Finally, we  discuss  and compare our findings with the published literature reports.
\end{abstract}


}]

%%include \doinum{number}for the DOI number in the header
%%include \volnum{number} for the volume number in the header
%%include \year{yyyy} for  year of publication in the header
%%include \pgrange{num--num} page range of article in the header
%%include \artcitid{num} for the article citation id
%%include \lp to print last page of the article
%%include \setcounter{page}{pagenum} for the exact starting page of the article

\doinum{XXXX}
\artcitid{\#\#\#\#}
\volnum{000}
\year{0000}
\pgrange{1--}
\setcounter{page}{1}
\lp{1}

\section{Introduction}\label{1}

The low mass X-ray binary (LMXB) GX 3+1 was first detected during an \textit{Aerobee}-rocket flight on June 16, 1964 (Bowyer et al., 1965). LMXBs are binary systems which are older than $\sim$ 10$^9$ years with their companion stars having mass $\leq$1 $M_\odot$ (Bhattacharya $\&$ van den Heuvel, 1991; Verbunt, 1993). Ever since the discovery of LMXBs, which have neutron stars as compact objects, short thermonuclear bursts have been reported (Grindlay et al., 1976; Belian et al., 1976; Strohmayer $\&$ Bildsten, 2006; Galloway et al., 2020). X-ray bursts in LMXBs are marked as rapid rise in the photon count in the time scale of secondsfollowed by an exponential decay (Galloway et al., 2008). These are the nuclear runaway phenomena, which are caused by the accreted material from the companion falling onto the NS surface via Roche-lobe overflow mechanism. Pure or mixed hydrogen burns upto a critical density, beyond which bursts bright in X-rays occur locally - see reviews by Lewin et al. (1993), Strohmayer $\&$ Bildsten, (2006) and Bhattacharyya, (2010). Usually the decay times vary between  $\sim$ 10-100 s as reported by Lewin, (1977); Hoffman et al., (1978); Lewin et al., (1993); Galloway et al. (2008). During spectral modelling, Type-1 bursts are normally described with blackbody models.
 
GX 3+1 is a persistently bright X-ray binary source. After eight years of its discovery,  it was first detected with Type-1 short bursts, which indicated that the compact object at the center certainly has to be an NS (Makishima et al., 1983). As reported from the All Sky Monitor (ASM) observation (Levine et al., 1996), the average X-ray intensity of GX 3+1 is  slowly varying over a time scale of months to years by a factor of $\sim$ 2 with high X-ray luminosity (10$^{37}$-10$^{38}$ ergs s$^{-1}$) and non-periodic behaviour (Mondal et al., 2019).

Being a luminous, persistent source GX 3+1 has been observed by many major X-ray missions such as \textit{Ginga} (Asai et al., 1993), \textit{EXOSAT} (Schulz et al.,1989), \textit{RXTE} (Bradt et al., 1993; Kuulkers $\&$ van derKlis, 2000), \textit{BeppoSAX} (den Hartog et al., 2003), \textit{INTEGRAL} (Chenevez et al., 2006; Paizis et al., 2006), \textit{Chandra} (van den Berg et al., 2014), \textit{XMM-Newton} (Pintore et al., 2015), \textit{NuSTAR} (Mondal et al., 2019). This present work is based on the observation of GX 3+1 by mission \textit{AstroSat}.

 LMXBs are mainly grouped as Z or Atoll sources, based on the characteristic shapes they trace on their color-color diagrams (CCD). Atoll sources display mainly two tracks in hardness-intensity diagrams: the banana and island states. GX 3+1 has been reported to be a bright atoll source having a soft spectrum typically $\sim$ 2-10 keV (Mondal et al., 2019). Ever since its discovery this source has been always found in banana state. Two branch structures were detected in its hardness intensity diagram (HID) which were identified as lower and upper banana states in the report by Asai et al. (1993). No island state has been observed so far from GX 3+1 and  no kHz Quasi-periodic oscillations (QPOs) are detected. (Homan et al., 1998; Oosterbroek et al., 2001; Chenevez et al., 2006, Mondal et al., 2019).

Historically, GX 3+1 has shown doubly peaked bursts in 9-22 keV energy band reported by Makishima et al. (1983) which was the first report of Type-1 bursts from GX 3+1. Following this, GX 3+1 was found to be a very active X-ray burster and bursts  were studied by Asai et al. (1993); Pavlinsky et al. (1994); Molkov et al. (1999). However, the first investigation of a double peak photospheric radius expansion burst (PRE) was carried out by Kuulkers $\&$ van der Klis (2000) using \textit{RXTE} data. They had estimated the source distance to be between 4.2 - 6.4 kpc for an uncertainity of 30 $\%$. A total of 61 bursts from GX 3+1 was reported by den Hartog et al. (2003) which were observed in a high state of the source. However, an exceptional superburst was reported by Kuulkers (2002) with \textit{RXTE/ASM} data, which had a decay time of $\sim$ 1.6 hours. An unusual intermediate burst with a duration of $\sim$ 30 min was detected on 2004 by \textit{INTEGRAL/JEM-X}, which was analysed by Chenevez et al. (2006). The same data set was re-evaluated by Alizai et al. (2020) using different spectral models. Thus, we find that no burst phenomenon has been reported for this source since 2004.

Thermonuclear X-ray bursts from GX 3+1 have never been reported earlier using AstroSat data. However, type-1 themonuclear bursts has been detected by AstroSat previously from various LMXBs (Bhattacharyya et al., 2018; Beri et al., 2019; Devasia et al., 2021; Bhulla et al., 2020; Roy et al., 2021, Kashyap et al., 2022). PRE burst events for the LMXB 4U 1636-536 have been reported earlier using AstroSat data by Beri et al. (2019) and Roy et al. (2021).

The spectral fitting of NS bursts involve modelling with a single color blackbody component (Hoffman et al., 1977; Beri et al., 2019; Bhulla et al., 2020). The pre-burst spectrum of several LMXBs are found to be well fitted with a thermally Comptonized model, which describes the powerlaw behaviour of the energy (Pintore et al., 2015; Verdhan Chauhan et al., 2017; Bhattacharyya et al., 2018; Chen et al., 2019). The parameter values of the black-body model could be used to estimate the temperature near the NS surface kT$_{\rm bb}$ $>$ 1 keV (Seifina \& Titarchuk, 2012)  and the physical radius of the NS. The model parameters evolve with time in rapid events like bursts, so it is needful to divide  bursts into short time intervals and fit the spectra of each interval, from the rise to the decay of the burst, which is the conventional method of time-resolved spectroscopy (Lewin et al., 1993; Swank et al., 1977). The source's persistent pre-burst spectra is assumed to be unevolved during the burst and used to serve as the background spectrum for the burst spectra. This is the conventional method of burst spectral analysis. Since type-1 bursts are  highly luminous than the source's average photon count rate, so it is acceptable to consider modelling of the  burst spectra distinguishably assuming the burst doesn't influence the background persistent emission. But reports have showed that bursts can modify the persistent spectra (Chen et al., 2012; in’t Zand et al., 2013; Degenaar et al., 2018; Keek et al., 2018). Worpel et al. (2013, 2015) introduced the ``\textit{f}$_a$'' (which is a scaling factor supplied to the persistent spectrum) method and reported the change in persistent flux due to the burst. When the ``\textit{f}$_a$''  is fixed at unity, the model follows the conventional method.  

The objectives of this paper are summarised in the following: (i) reporting the detection of a type I X-ray burst from GX 3+1 with \textit{AstroSat}, (ii) reporting the detection of the double-peak feature in the burst at higher energies, and, (iii) presenting spectral analysis of the burst and reporting the physical parameters such as the blackbody temperature, the blackbody flux and the radius of the NS photosphere.

The organization of this paper is as follows- Section 2 discusses the observational details and mentions all the data reduction techniques briefly. Section 3 presents the light curve, the hardness ratio of the photon count rate, the hardness-intensity diagram (HID), an energy resolved burst light curve and QDP modelling of the burst at each energy band. We have performed a joint fitting of the persistent pre-burst spectrum, using SXT and LAXPC 20 data and obtained the parameter values for a thermally Comptonized component, elaborated in Section 4.1. We have adopted the conventional approach to fit model the burst, elaborated  in Section 4.2. All the results are discussed and compared to highlight the consistencies with earlier published reports in Section 5.

\section{Observations and Data Reduction }\label{sec2}
\textit{AstroSat} is India's first multi-wavelength astronomy mission satellite launched on September 28, 2015. For dedicated studies on X-ray astronomy, \textit{AstroSat} has been provided with payloads Soft X-ray focusing Telescope (SXT) and three Large Area X-Ray Proportional Counters (LAXPC) named as LAXPC 10, LAXPC 20 $\&$ LAXPC 30. \textit{AstroSat} observed GX 3+1 in an Announcement of Opportunity (AO) observation conducted in April 29-30, 2018 (Obs ID:A04$\_$122T01$\_$9000002064). The observation was done for the Right Ascension (RA) $\alpha$ =  266.983329 and Declination point (DEC) $\delta$ = -26.56361 in International Celestial Reference System (ICRS), for the celestial co-ordinates of GX 3+1. The Observation by \textit{AstroSat} is marked on the MAXI long time light curve in 2-20 keV shown in Fig. 1. The source was showing a persistent behaviour during the \textit{AstroSat} observation, as it is evident from the MAXI light curve.

\begin{figure}[!t]
\centering
\includegraphics[width=\columnwidth]{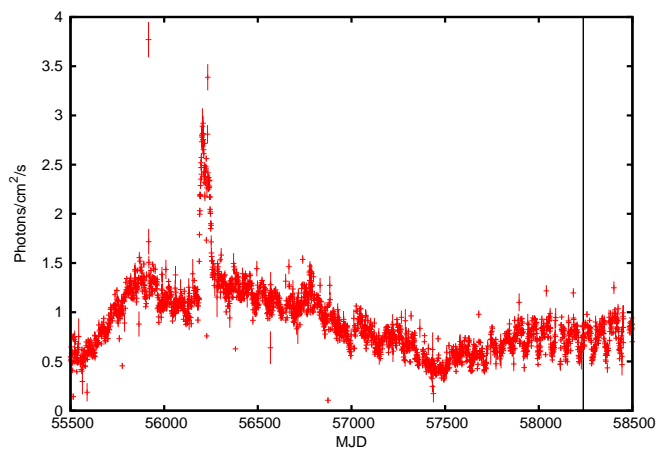}
\caption{The MAXI long time Light curve for GX 3+1 in the energy band 2-20 keV. The vertical line marks the \textit{AstroSat} observation date.} 
\label{maxi}
\end{figure}

\subsection{\textit{AstroSat}-SXT}
\label{sxtsec}
The SXT (Singh et al., 2016, 2017) has an operational energy band of 0.3-8.0 keV (Singh et al., 2017; Bhattacharyya et al., 2021). The SXT pipeline software\footnote{http://astrosat-ssc.iucaa.in/?q=sxtData} (version: AS1SXTLevel2-1.4b) is used to generate the Level 2 data of 9 orbits using the Photon Counting (PC) mode for Level 1 data. The SXT event merger script is used to merge the data for different orbits to produce the clean event file. An encircled region of radius 13.5 arcmin in physical coordinates is extracted which comprises $\sim$ 96$\%$ of the source photons, for the generation of source spectrum. This extraction is done using the standard tools of XSELECT V2.4g. The light curve of minimum allowed time bin for SXT instrument i.e 2.3775 s, is thus obtained from this region file. The Auxillary Response File (ARF) is generated by using sxtARFModule{\footnote{www.tifr.res.in/$\sim$astrosat$\_$sxt} tool of the ARF on-axis (version 20190608) provided by the SXT instrument team. The SXT spectrum file is used with a blank sky background spectrum provided by the SXT instrument team during our model fitting. No pile up is observed for the source. The energy band during the generation of the SXT spectrum file is kept the default 0.3-8.0 keV (Singh et al., 2017; Bhattacharyya et al., 2021).

\subsection{\textit{AstroSat}-LAXPC}
LAXPC counters have been designed to detect  X-ray photons of energies 3.0-80 keV (Yadav et al., 2016; Antia et al., 2021). With a total of three proportional counters as mentioned before, LAXPC has a total effective area of $\sim$ 8000 cm$^2$. Each of the three counters work independently to record  photons with a time resolution of roughly 10 $\mu$s. This makes LAXPC able to observe fast variability like thermonuclear bursts. We use the data only from LAXPC 20 because LAXPC 10 has been found displaying instability in its response  and LAXPC 30 is excluded as it has been officially shut down, at the time of observation of GX 3+1.
 
We have generated the Level 2 data of 9 orbits from LAXPC 20 using the Event Analysis (EA) mode. EA mode data contains the information about the time, anodeID and Pulse Height Amplitude (PHA) for each event. The processing of the LAXPC data is done using individual routine software  LaxpcSoft\footnote{http://astrosat-ssc.iucaa.in/?q=laxpcData} version: May 19, 2018. We have only used the layer 1 (top layer) of the LAXPC 20 counter for the temporal and spectral analysis except for the full time light curve, which is generated using all the layers of the same counter.

\section{Temporal Analysis}\label{temp}
 The LAXPC 20 light curve is generated for the good time intervals only. The background is subtracted for a default $2\%$ systematic error. A nearly persistent photon count rate of $\sim$ 1000 counts/s is observed from the light curve. As we cross about 60 ks from the observational start time, the burst feature is prominent (Fig. 2) with its photon count rate ($\sim$ 4000 counts/s) almost four times the non-burst rate. 

\begin{figure}[!t]
\centering
\includegraphics[scale=0.33,angle=270]{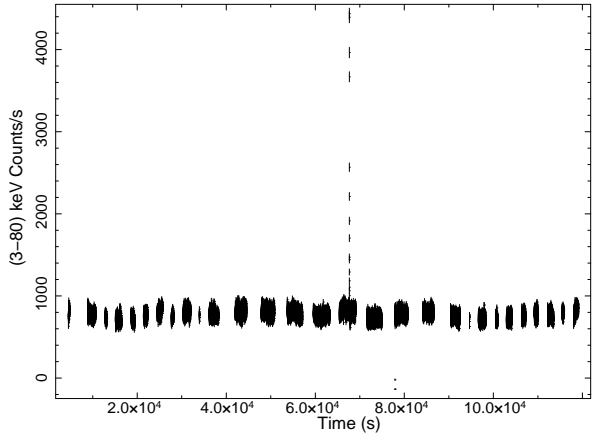}
\caption{\textit{AstroSat}/LAXPC counter 20 light curve for $\sim$ 100 ks data in the 3.0-80 keV energy band for a bin size of 1.0 seconds. A burst is observed at nearly 60 ks from the start of the observation with a sudden increment in the count rate by approximately 4 folds of the source's persistent count rate.} 
\label{lcfull}
\end{figure}

The Fig. 3 shows the detected photons in LAXPC 20 layer 1 in energy bands 3-6 keV and 6-12 keV, with their respective light curves plotted simultaneously and the hardness ratio (HR) (6-12 keV/3-6 keV) of the two light curves in the bottom panel , with their corresponding hardness ratio shown at the bottom panel. The bin size is selected as 0.2 s. We have used xronos v5.22 is used to generate this multiplot. We can thus infer that during the entire observation period,  the source is detected in a soft state since the average hardness ratio was $\sim$ 0.6-0.7 during the entire observation. But during the occurrence of the burst, we notice the ratio increases, crossing 1.0, followed a sharp drop in the hardness and a subsequent rise. As the burst decays, the ratio drops to the persistent level.

The HID is plotted as a function of source's intensity (Fig. 4). For an HR (6-12 keV)/(3-6 keV) with a binning size of 60 seconds, we observe a positive correlation of the HR with source intensity in the energy band 3-12 keV. A positive correlation between hardness and intensity is also reported by Mondal et al. (2019) which indicates the characteristic soft banana state. 

\subsection{Search for Burst Oscillations}

We have investigated the power density spectrum (PDS) to detect burst oscillations using data from LAXPC 20 all layers. To generate the PDS, we have used the Laxpcsoftware task ``\textit{laxpc$\_$find$\_$freqlag}'' which outputs the power spectra as a function of frequency. These PDS are generated using a burst GTI file of 16 s exposure. We have also used the GTI file for the entire orbit data from LAXPC 20 all layers in which the burst has been detected and plotted the respective PDS upto 1000 Hz. We have also parallely used the ftool \textit{powspec} 1.0 (XRONOS v 6.0) to generate PDS with binning of 0.005 s, 0.004 s, 0.003 s and 0.002 s. However, we could not detect any signal of oscillations.

\begin{figure}[!t]
\centering
\includegraphics[scale=0.33,angle=270]{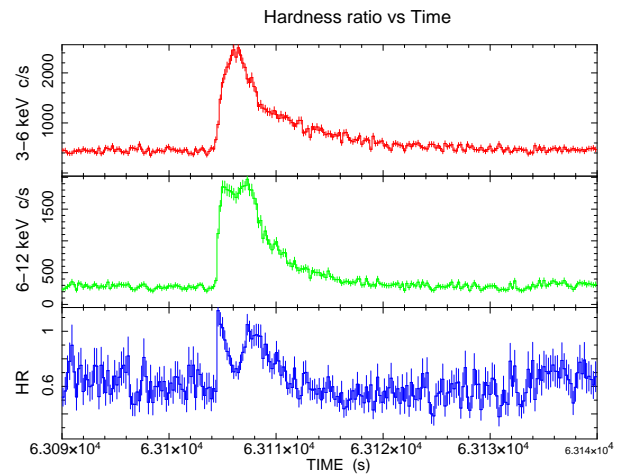}
\caption{The upper and middle panels show source LAXPC 20 layer 1 light curves in the energy band 3-6 keV and 6-12 keV respectively. The bottom panel shows the hardness ratio HR = $\big(\frac{6-12{\:\rm keV}}{3-6{\:\rm keV}}\big)$. The binning size is 0.2 seconds. An increase in HR is seen during the burst, crossing 1.}
\label{HR}
\end{figure}
\begin{figure}[!t]
\centering
\includegraphics[scale=0.33,angle=270]{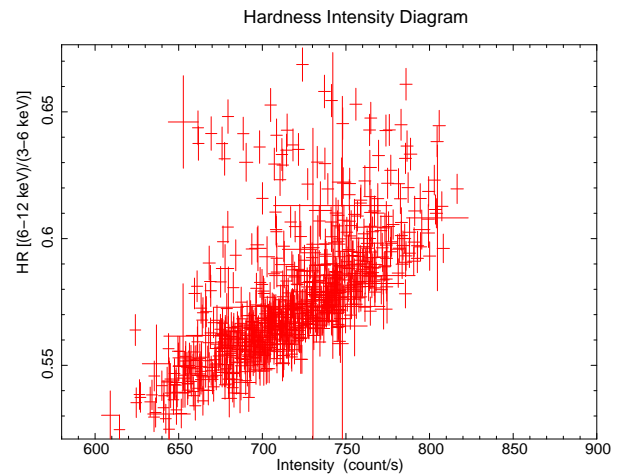}
\caption{The hardness-intensity diagram of GX 3+1. The hardness ratio (HR) is taken as the ratio of the count rates in energy bands 6-12 keV and 3-6 keV. The HR shows a positive correlation with the intensity. The plot is generated for a bin time of 60.0 s.}
\label{HIDGXlax}
\end{figure}

\subsection{Energy-resolved burst profile}
\label{burst_evol}

To resolve the entire burst into separate energy bands, we generate four burst light curves in narrow energy bands: 3-5 keV, 5-8 keV, 8-12 keV and 12-20 keV. All of these light curves have a binning size of 0.16 seconds. From Fig. 5, we observe the burst has the highest photon count rate in 5-8 keV band, crossing  2000 counts/s. This is followed by the light curves in 3-5 keV, 8-12 keV and 12-20 keV band in a descending order of photon count rate. The double peak feature is observed in the 8-12 keV and 12-20 keV bands. These burst light curves are generated after subtracting the background light curves, using a default systematic error of 2$\%$ for the LAXPC instrument and no scaling factors are multiplied.
 \begin{table*}[!t]
\centering

\caption{The duration of rise, peak to persistent count ratio, decay duration and flux at the burst peak estimated for four narrow energy bands of the burst. The count ratio is the  approximate ratio of peak count rate and the persistent count rate, obtained for each \textit{Burs} model}
\label{tab:qdp}
{\begin{tabular}{|c|c|c|c|c|}
\hline
Energy band (keV) & Duration of rise(s) & Count ratio & Decay duration(s)& Peak Flux (10$^{-9}$ ergs s$^{-1}$ cm$^{-2}$)\\
\hline
3 - 5 & 0.845, 1.05 & $\sim$ 3.18, 5.12 & 1.7, 6.81 & 9.763\\ 
5 - 8 &  0.72, 3.15& $\sim$ 6.96, 4.24 & 5.98, 1.56 & 11.471\\
8 - 12 & 0.52, 1.616& $\sim$ 8.03, 7.58 & 2.87, 1.89 & 8.430 \\
12 - 20  & 0.4, 1.97& $\sim$ 10.43, 8.69 & 1.22, 1.59 & 5.270\\
\hline

\end{tabular}}

\end{table*}
We have performed a measurement of the exponential decay times of the burst in four energy bands by fitting the corresponding light curves with a model combination of a constant added to QDP \textit{Burs} model\footnote{https://heasarc.gsfc.nasa.gov/ftools/others/qdp/qdp.html} (Devasia et al., 2021). To achieve a better fitting, we include two \textit{Burs} models. We show the model fits for four burst profiles  in their respective energy bands in Fig. 6 and report the obtained rise time, the approximate ratio of the burst peak count rate and the persistent count rate and the decay time, in Table. 1. We find that the peak to persistent count rate ratio is higher for higher energy bands, as evident from Fig. 6(a-d). QDP \textit{Burs} modelling of burst profile has been done previously by Marino et al. (2019), Beri et al. (2019) and Hu et al. (2020) to obtain their timing properties. The peak flux values listed in Table. 1 have been obtained by using the XSPEC command \textit{flux} on the spectral files of exposure $\sim$ 3 s near the peak of each of the energy resolved burst light curve.

\begin{figure}[!t]
\centering
\includegraphics[width=\columnwidth]{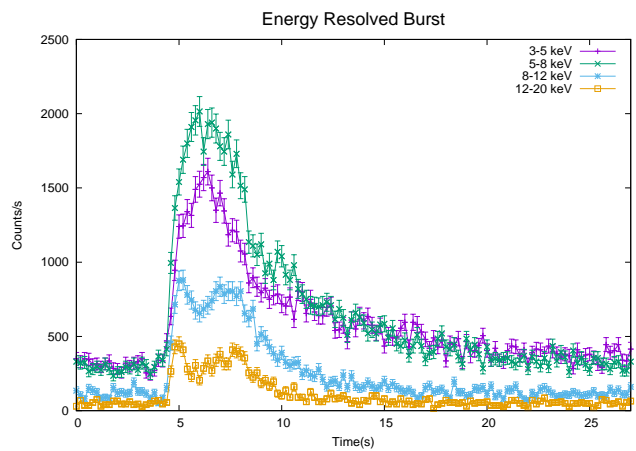}
\caption{The plot shows the thermonuclear X-ray burst observed in different energy bands. The X-ray burst decays rapidly as energy approaches to 20 keV. The highest photon count is observed for the energy band 5-8 keV. All the light curves have a binning size of 0.2 s.}
\label{bursten}
\end{figure}

\begin{figure*}[!t]
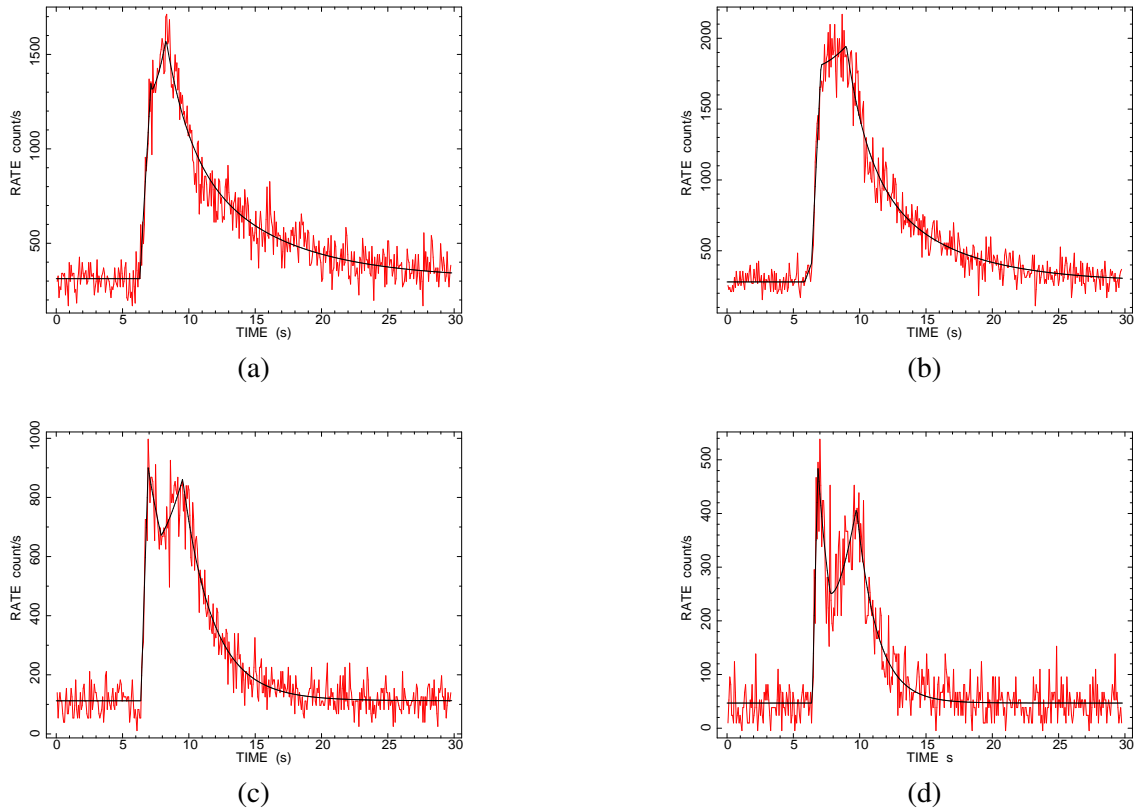

\begin{subfigure}{0.5\textwidth}
\centering
\includegraphics[angle=-90,width=0.8\textwidth]{fig6_a.eps}
\caption{}
\end{subfigure}
\begin{subfigure}{0.5\textwidth}
\centering
\includegraphics[angle=-90,width=0.8\textwidth]{fig6_b.eps}
\caption{}
\end{subfigure}
\begin{subfigure}{0.5\textwidth}
\centering
\includegraphics[angle=-90,width=0.8\textwidth]{fig6_c.eps}
\caption{}
\end{subfigure}
\begin{subfigure}{0.5\textwidth}
\centering
\includegraphics[angle=-90,width=0.8\textwidth]{fig6_d.eps}
\caption{}
\end{subfigure}
\caption{QDP modelling of the Type-1 burst. Fig.(a) shows modelling of the burst in 3-5 keV energy range, Fig.(b), in 5-8 keV energy range, Fig.(c) in 8-12 keV energy range and Fig.(d), models the burst light curve in 12-20 keV energy range. At higher energies, the double peaked feature is seen more distinctively.}
\label{qdp}
\end{figure*}

\section{Spectral Analysis}

To perform the spectral fitting, we have used the SXT spectrum file, mentioned in Section 2.1. We have considered only  layer 1 data from LAXPC 20 unit to obtain non-burst or pre-burst spectrum and the burst spectrum. We consider two broad regions of data - the pre-burst region and the burst region.
\subsection{Pre-burst Analysis}\label{sapreb}
\begin{table*}
	\centering
		\caption{The best-fit parameter values for the joint pre-burst spectral modelling of SXT and LAXPC 20 layer 1 data with a thermally Comptonized multicolour black-body Constant$\times$TBABS$\times$NTHCOMP. Errors are at 90$\%$ confidence range for each parameter}
	\label{tab:preburst}
	\begin{tabular}{ c c c c } % four columns, alignment for each
		\hline
		Component & Parameter(Unit) & LAXPC 20 & SXT\\
		\hline 
		\hline
		TBABS & N$_{\rm H}$(10$^{22}$ cm$^{-2}$) & 1.4 (frozen) & 1.4 (frozen) \\\\ 
		\hline
		Constant &  & 1.0 (frozen) & 1.18$^{+0.04}_{-0.04}$ \\                 
		\hline
		NTHCOMP & $\Gamma^\alpha$ & 3.5$^{+0.4}_{-0.3}$ & 3.5$^{+0.4}_{-0.3}$\\\\
		  & kT$_{\rm e}^\beta$(keV) & 4.46 $^{+2.55}_{-0.88}$ & 4.46$^{+2.55}_{-0.88}$ \\\\
		  & kT$_{\rm nth}^\gamma$(keV) & 1.82$^{+0.07}_{-0.07}$ & 1.81$^{+0.07}_{-0.07}$ \\\\
		  & N$_{\rm nth}^\delta$ & 0.73$^{+0.04}_{-0.04}$ & 0.73$^{+0.04}_{-0.04}$\\\\
		\hline
		Gain & Slope & - & (1.0)\\\\
		& Offset (E-02) & -  & 5.73$^{+0.006}_{-0.007}$ \\\\
		\hline
		  &$\chi^2$/d.o.f &  1.08 & 1.08 \\\\
		\hline  
		  & $f_{\rm}^\eta$(10$^{-9}$ ergs s$^{-1}$ cm$^{-2}$) & 4.46$^{+0.1}_{-0.1}$ & 6.41$^{+0.1}_{-0.1}$ \\\\
		\hline
		
	\end{tabular}
	
     {\raggedright 
     $^\alpha$ The powerlaw photon index of NTHCOMP. $^\beta$ Electron temperature of the Corona. $^\gamma$ Seed photon temperature of the disk. 
     
     $^\delta$ Normalisation of NTHCOMP. $^\eta$ Unabsorbed flux in the energy 4.0-16.0 keV for LAXPC 20 and 0.6-7.0 keV for SXT data. }
    
	\end{table*}
	
The pre-burst energy spectrum of the source is obtained from the combined analysis of the SXT and LAXPC 20 layer 1 spectral data of $\sim$ 841 s, which is modelled by a thermally Comptonized component NTHCOMP (Zdziarski et al., 1996; Życki et al., 1999) in XSPEC v12.10.1f (Arnaud, 1996). We choose an energy range with a lower limit at 0.6 keV because of the uncertainties in the response at lower energies and higher limit at 16 keV, since the spectra is found to be background dominated at higher energies. The XSPEC routine TBABS (Wilms et al., 2000) is used for taking the interstellar medium (ISM) absorption into account. We freeze the neutral hydrogen column density (N$_{\rm H}$)  at 1.4 $\times$ 10$^{22}$ cm$^{-2}$ (Fig. 7) which is close to the value $\sim$ 1.5 $\times$ 10$^{22}$ cm$^{-2}$ reported by Morrison $\&$ McCammon (1983). We select the input type 1 which considers that the seed photons are supplied to the corona by the accretion disk.
 
 Table. 2 shows all the parameter values achieved for the best fit. The LAXPC 20 layer 1 spectrum and the SXT spectrum are obtained for energy bands 4.0 - 16.0 keV and 0.6 - 7.0 keV respectively. The SXT spectrum's response file is provided a gain correction, with its slope frozen to 1.0 and its best fit offset obtained $\sim$ 0.047$^{{+0.008}}_{-0.012}$. A systematic error of 3$\%$ is considered to account for the uncertainties in the response calibration\footnote{https://www.tifr.res.in/~astrosat$\_$sxt/dataana$\_$up/readme$\_$sxt$\_$arf$\_$data$\_$analysis.txt} (Maqbool et al., 2019; Mudambi et al., 2020).
   
   \begin{figure}[!t]
 \centering
\includegraphics[scale=0.33, angle=270]{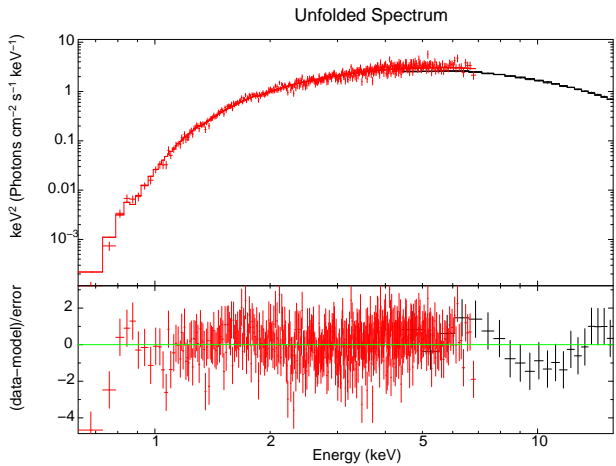}
\caption{The unfolded spectra from \textit{AstroSat} for GX 3+1 (SXT spectrum in red and LAXPC 20 layer 1 spectrum in black). The top panel shows the spectra fitted with the model Constant$\times$TBABS$\times$NTHCOMP . The energy band used is 0.6 - 16.0 keV. The bottom panel shows the ratio of data and model.}
\label{prebdbb}
\end{figure}
   While fitting, the photon index $\Gamma$ is found to settle at $\sim 3.5 \pm 0.35$, which is in agreement with the reported values from Chenevez et al. (2006); Pintore et al. (2015); Ludlam et al. (2019). The electron temperature kT$_{\rm e}$   is obtained $\sim$ 4.46 $^{{+2.55}}_{-0.88}$ keV and the seed photon temperature kT$_{\rm nth}$ is achieved $\sim$ 1.82 $^{{+0.07}}_{-0.07}$ keV. The low value of the electron temperature indicates to the soft state of the source, which is in agreement with our obtained HID shown in Fig. 4. The fit attains a chi-squared value of 471.76 for 437 degrees of freedom, which is a good fitting. The SXT spectrum is rescaled by a constant factor $\sim$ 1.18 $^{{+0.04}}_{-0.04}$, while keeping the same constant fixed to 1.0 for LAXPC 20 layer 1 spectrum. 

  The unabsorbed flux is found to be slightly more for the SXT spectra in comparison to the LAXPC 20 layer 1 spectra (4.46 and 6.41 in units of 10$^{-9}$ ergs s$^{-1}$ cm$^{-2}$, for LAXPC and SXT respectively), obtained using the convolution model CFLUX. These values are in agreement with the persistent flux reported by Pintore et al. (2015); Ludlam et al. (2019).
  
We have also attempted to add a DISKBB model to the NTHCOMP, but it has resulted in an overestimation of the errors in the fit. We have also tried to fit the pre-burst spectra by selecting the input type=0 so as to remodel the spectrum with blackbody seed photons. This has resulted in a poor fit, with a reduced chi-squared value rising to $\sim$ 1.21. However, adding a BBODY model to the NTHCOMP slightly lowered the reduced chi-squared value to $\sim$ 1.18, but the obtained normalization is found very small and its error limits couldn't be constrained.
 
 \subsection{Burst Analysis}\label{saburst}
We investigate the evolution of the free parameters during the dominant phase of the burst, by systematically dividing the burst exposure into a total of five time intervals- T1, T2, T3, T4 and T5, shown in Fig. 8. The exposure of the different time intervals are T1 = 0.8 s, T2 = 1.0 s, T3 = 0.9 s, T4 = 0.6 s and T5 = 2.4 s. We obtained the spectra for each of these five time bins from the top layer or layer 1 of the LAXPC 20. The energy band selected during XSPEC fitting  is 4 - 16 keV. Harder energies more than 16 keV are ignored as the burst is observed predominantly present at softer energies (see Fig. 5). Energies lesser than 4.0 keV are ignored to avoid uncertainties in the response from LAXPC. As the model TBABS $\times$ NTHCOMP has described the pre-burst spectra successfully, we proceed to investigate the burst by adding a blackbody model (Degenaar et al., 2016) BBODYRAD to pre-burst model (Galloway et al., 2008; Beri et al., 2016). The NTHCOMP is fed with pre-burst parameter values which are kept frozen and only the blackbody parameters are set free. This conventional approach is adopted to model all the five burst spectra, shown in Fig. 9. A systematic error of 2\% is considered to account for the uncertainties in the spectral response of LAXPC following the works of Antia et al. (2017); Misra et al. (2017); Sreehari et al. (2019, 2020); Kashyap et al. (2022); Majumder et al. (2022).

The Table. 3 reports the blackbody temperature and the normalisation constant achieved during the fitting. We notice that the temperature kT$_{\rm bb}$ reaches a local minimum of 1.61 $\pm$ 0.08 keV corresponding to which the Norm$_{\rm bb}$ reaches the maximum value = 496.05$^{{+138.64}}_{-106.98}$. During T5, which is the decay phase of the burst, the blackbody temperature rises again following a drop in the normalisation value. We could also observe that the unabsorbed flux is nearly constant from time bin T2 to T4 (the time interval during which the burst has shown the double peak at higher energies). We highlight that at T2, the Norm$_{\rm bb}$  is obtained with an almost six fold increase from T1. The normalisation values are used to estimate the radius using the relation\footnote{https://heasarc.gsfc.nasa.gov/xanadu/xspec/manual/node137.html} Norm$_{\rm bb}$= R$^2$/D$^2_{10 \rm kpc}$ and we achieve the maximum value of the photospheric radius 13.58$^{{+1.89}}_{-1.46}$ km which agrees with the inner disk radius reported by Mondal et al. (2019) using \textit{NuSTAR} data, in which they have also mentioned an upper limit of $\leq$ 13 km to the NS radius. To showcase all these parametric changes involved during the burst span, we have plotted the physical parameters vs time with t=0 as the burst rise in Fig. 10. We have not used any color correction factor to derive the radius and the burst temperature. The radii values are estimated for a source distance of  6.1 kpc (Kuulkers $\&$ van der Klis, 2000) reported for a helium rich photosphere.

\begin{figure}[!t]
\centering
\includegraphics[scale=0.33, angle=-90]{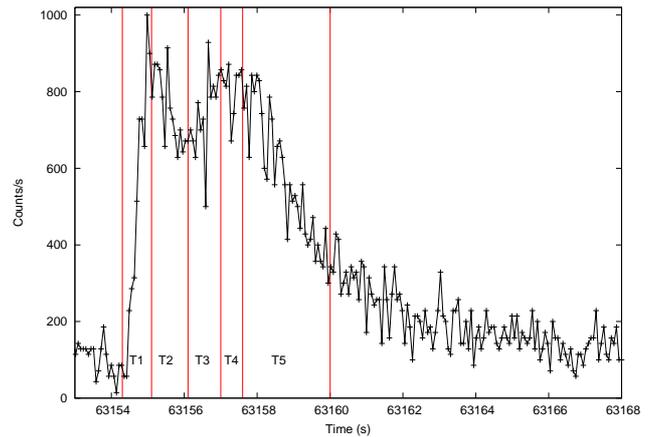}
\caption {The time bins selected for spectroscopy. The burst light curve shown here is in 8-12 keV energy band.}
\label{burstbin}
\end{figure}

\begin{figure*}[!t]
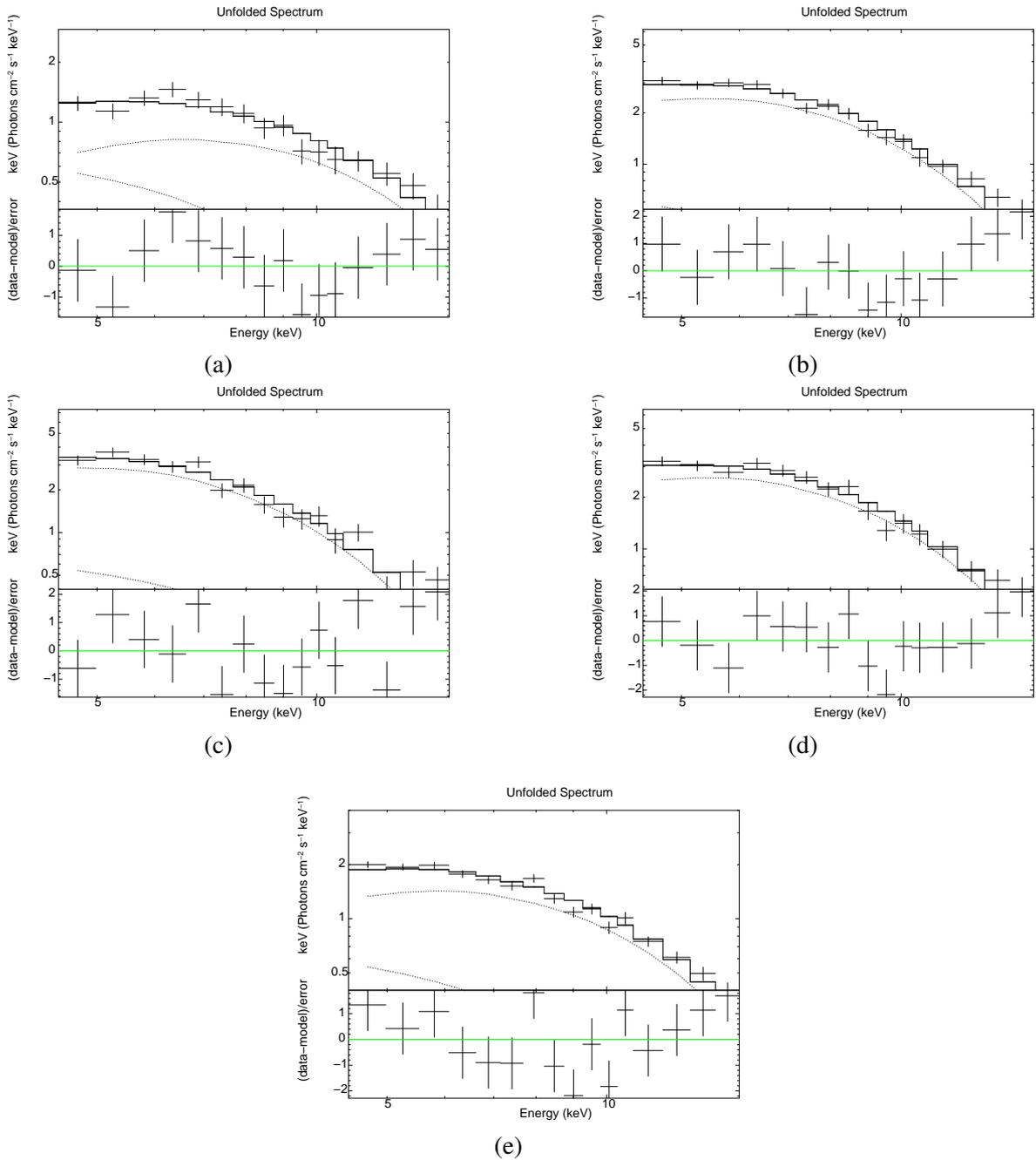

\begin{subfigure}{0.5\textwidth}
\centering
\includegraphics[angle=-90,width=0.8\textwidth]{fig9_a.eps}
\caption{}
\end{subfigure}
\begin{subfigure}{0.5\textwidth}
\centering
\includegraphics[angle=-90,width=0.8\textwidth]{fig9_b.eps}
\caption{}
\end{subfigure}
\begin{subfigure}{0.5\textwidth}
\centering
\includegraphics[angle=-90,width=0.8\textwidth]{fig9_c.eps}
\caption{}
\end{subfigure}
\begin{subfigure}{0.5\textwidth}
\centering
\includegraphics[angle=-90,width=0.8\textwidth]{fig9_d.eps}
\caption{}
\end{subfigure}
\begin{center}
\begin{subfigure}{0.5\textwidth}
\centering
\includegraphics[angle=-90,width=0.8\textwidth]{fig9_e.eps}
\caption{}
\end{subfigure}
\end{center}
\caption{Time resolved spectroscopy of the burst using 5 time bins. We modelled the spectra during rise T1 shown by Fig.(a), Fig.(b), (c), (d) and (e) are the unfolded modelled spectra corresponding to the time bins T2, T3, T4, and T5. Each of the spectra is fitted following the conventional method on TBABS $\times$ (NTHCOMP + BBODYRAD). NTHCOMP parameters are frozen at pre-burst values for all the spectra.}
\label{spectra}
\end{figure*}

\begin{table*}
\centering
    \caption{{The best-fit parameters of BBODYRAD achieved during time resolved spectroscopy of the burst. The model fitted is TBABS $\times$ (NTHCOMP + BBODYRAD). The parameter values of TBABS and NTHCOMP are kept frozen to the values achieved during modelling of the pre-burst emission.The energy band taken here for all the time bins is 4.0-16.0 keV. All the errors are at 90$\%$ confidence range}}
    \label{tab:burst}
{\begin{tabular}{ c c c c c c  }
 \hline
		 Parameter & T1  & T2 & T3 & T4 & T5\\
		\hline 
		\hline
    &&BBODYRAD\\
    \hline
    kT$_{\rm bb}^\alpha$ &2.30$_{-0.14}^{+0.16}$ &1.84$_{-0.06}^{+0.06}$ &1.61$_{-0.08}^{+0.08}$ &1.83$_{-0.07}^{+0.07}$ &2.0$_{-0.06}^{+0.06}$\\\\
    Norm$_{\rm bb}^\beta$ &46.82$_{-11.10}^{+14.13}$ &269.42$_{-38.3}^{+44.8}$ &496.05$_{-106.98}^{+138.64}$ &301.40$_{-50.51}^{+60.7}$ &124.78$_{-15.09}^{+17.18}$\\\\
    \hline
    Radius$^\gamma$ &4.16$_{-0.47}^{+0.60}$ &9.43$_{-0.75}^{+0.88}$ &13.58$_{-1.46}^{+1.89}$ & 10.6$_{-0.88}^{+1.06}$&6.81$_{-0.41}^{+0.46}$\\\\
    $\chi^2$/d.o.f&0.81 &1.16 &1.65 &1.12 &1.67\\\\
    Flux$^\delta$ & 1.12$_{-0.15}^{+0.15}$ & 2.58$_{-0.24}^{+0.24}$ & 2.54$_{-0.23}^{+0.23}$ & 2.76$_{-0.26}^{+0.26}$ & 1.67$_{-0.1}^{+0.1}$\\\\
    \hline
 
\end{tabular}}

{{\raggedright  
$^\alpha$ The blackbody temperature in keV units.

$^\beta$ Normalisation constant of BBODYRAD in given by Norm$_{\rm bb}$= R$^2$/D$^2_{10 \rm kpc}$, where R is the radius in km and D$^2_{10 \rm kpc}$ is the source distance in units of 10 kpc.

$\gamma$ Radius of the photosphere in units of km, estimated for a source distance of 6.1 kpc.

$^\delta$ Unabsorbed blackbody flux in 10$^{-8}$ ergs s$^{-1}$ cm$^{-2}$ units.

\par}}
\end{table*}

\begin{figure*}[!t]
        \centering
        \includegraphics[scale=0.6, angle=0]{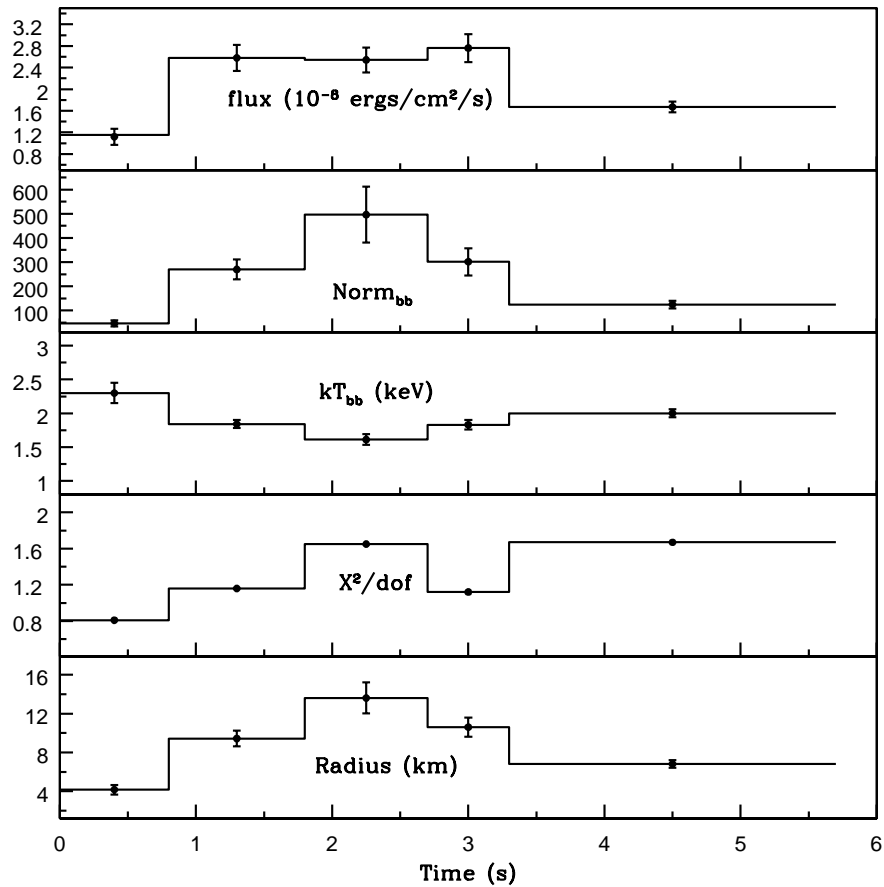}
\caption{A combined plot of the evolution of unabsorbed flux, the normalisation constant of blackbody model, blackbody temperature kT$_{bb}$, the model error and the estimated photospheric radius in km for a source distance of 6.1 kpc. The x-axis counts the time from the rise of the burst.}
\label{multi}
\end{figure*}

\section{ Discussions}\label{sec4}

In this work, we report the \textit{AstroSat} observation of the bright atoll source, the LMXB GX 3+1. The light curve obtained from LAXPC 20 instrument indicated the presence of a  thermonuclear burst feature of Type-1. Our temporal analysis revealed that the instrument SXT had observed the source simultaneously with the LAXPC 20 during the pre-burst stage till $\sim$ 2 minutes before the burst was detected by LAXPC 20. Since LAXPC is operational during both orbital day and night and SXT can only observe during orbital night, there is more exposure by LAXPC than SXT prior to the satellite entering the South Atlantic Anomaly (SAA) passage. Hence, unfortunately, SXT data is not available during the thermonuclear burst.\\

It's noteworthy that the large collective area of LAXPC has allowed us to generate an energy resolved burst profile and we have followed Beri et al. (2019) and Bhulla et al. (2020). A drop in count rate in the light curve of the X-ray burst is observed when the narrow energy bands become harder. Also, we found that the burst was brightest in 5-8 keV energy band with a count rate higher than the softest band 3-5 keV. From our QDP modelling, we find that the burst decays faster at higher energies, which indicates that the temperature is decreasing as the burst evolves. This trend is similar to the behaviour of the burst from 4U 1636-536, an atoll source, reported by Beri et al. (2019). We also noticed a double - peak feature in the burst at higher energies i.e 8-12 keV, 12-20 keV. This is a quick event $\sim$ 2 s within which the burst had showed the double peak, indicating a radius expansion phase (Watts $\&$ Maurer, 2007; Paczynski, 1983). It is necessary to highlight that double peak bursts are also observed for bursts categorized as non-PRE bursts (Regev $\&$ Livio, 1984), where the double peak feature is observed for low energies as well, theoretically modelled by Fujimoto et al. (1988); Melia $\&$ Zylstra (1992); Fisker et al. (2004); Bhattacharyya $\&$ Strohmayer (2006); Zhang et al. (2009); Lampe et al. (2016) and Bult et al. (2019).

We now discuss about our modelling of the pre-burst or persistent spectra. The seed photon temperature is found to be $<$ 1 keV in the literature reports of GX 3+1, but Pintore et al. (2015) has obtained values $\geq$ 1 keV for their two used models, both involving the thermally Comptonized NTHCOMP without fixing  the seed photon temperature with the disk blackbody temperature. In our case the persistent pre-burst spectrum is fitted by an NTHCOMP for a disk blackbody emission and without any explicit DISKBB (Mitsuda et al., 1984; Makishima et al., 1983). The $\rm kT_{nth}$ is obtained to be $>$ 1.8 keV. This is in close agreement to inner disk temperature reported by Mondal et al. (2019). We mention that the atoll source 4U 1728-34 is reported with seed photons $>$ 1.6 keV by Bhattacharyya et al. (2018) for an NTHCOMP model with disk blackbody as input (no explicit disk blackbody component is used). We also mention that we could not improve the fitting of the pre-burst spectra by adding DISKBB or BBODY models to NTHCOMP. Since we have used the NTHCOMP model only to model the pre-burst spectrum where blackbody seed photons were not the input, the Comptonizing layer must have primarily covered the disk (Bhattacharyya et al., 2018).

During the spectral modelling of the burst, we have used the model BBODYRAD which directly leads to estimate the radius of the source. We find that the photopshere shrinks and the radius reaches to a value $\sim$ 6.81 km during the decay of the burst. Bhattacharyya et al. (2018) systematically examined the relative differences in the blackbody normalization value achieved for both the conventional and the ``\textit{f}$_a$" method, and the latter has been found to provide lower normalisation values and hence a smaller NS radius. We have also tried adopting the $f_a$ methodology to model burst spectra, but the data quality is not good enough to constrain complex models. Referring to Fig. 10, we see that the photospheric radius increases and reaches a local maximum at the same time when the blackbody temperature kT$_{bb}$ reaches the  minimum value as it could seen at the top plot. From T2 to almost T4, which describes the PRE episode, the unabsorbed X-ray flux is found to remain approximately constant at $\sim$ (2.58-2.76 in 10$^{-8}$ ergs s$^{-1}$ cm$^{-2}$), which is expected for a PRE burst (Strohmayer $\&$ Bildsten, 2006; Galloway et al., 2008). During the double-peak event the source contracts and the photosphere expands thus the flux and hence the luminosity is expected to remain almost constant. This is derived from the understanding that luminosity is directly dependent to both the source temperature and the radius. The expansion of the photosphere leads to a drop in photon temperature and thus X-ray count rate drops (Tawara et al., 1984). Studies performed using the BeppoSAX data  by Beri et al. (2016) indicated the presence of high temperatures ($\sim$ 3 keV) for X-ray bursts showing double-peaked profiles at higher energies, however, we don't find the presence of such high temperatures during the touchdown period. Since we have used a frozen pre-burst model as a background to the burst spectra, hence we are unable to investigate the effect of burst on the persistent emissions during the PRE event which was previously carried out for a non-PRE burst by Bhattacharyya et al. (2018).

It is worthy to mention that during the years 1983-2006, most of the reports of bursts from GX 3+1 had been published after which no burst event is reported for almost more than a decade. As mentioned in our Introduction (Sec. 1.), the first measurement of the radius of the NS was carried out by Kuulkers $\&$ van der Klis (2000) and the authors have reported a radius of be 4.5 $\pm$ 0.3 km during the non-burst period. This is consistent with the radius value derived by us during T1 (rise of the burst, see Fig. 10) $\sim$ 4.16 $\pm$ 0.53 km. A single blackbody approximation on burst spectrum modelled by Molkov et al. (1999) had reported the NS radius to be 7.2 $\pm$ 1.2 km for a considered source distance of 8.5 kpc. If we consider a source distance of 8.5 kpc, we obtain the touchdown radius to be $\sim$ 9.5 km. To add, the estimation of source distance could be well derived during a PRE, and ever since Kuulkers $\&$ van der Klis (2000) had reported it, the source distance hasn't been calculated with a greater precision. Since we could detect a PRE event, we attempt to estimate the source distance. The peak luminosity during the PRE event is considered to reach the source's Eddington luminosity, as the NS surface is lifted up. We use this opportunity to  estimate the source distance by using the peak flux obtained by us, from Table. 3. The peak photon flux ($F_{\rm b}$) is related to the Eddington luminosity ($L_{\rm Edd}$) by a linear equation, which is the modified Stefan-Boltzmann law for an LMXB, given by Lewin et al. (1993): 
\begin{equation}\label{first}
L_{\rm Edd}= 4\pi d^2\xi_{\rm b}F_{\rm b}
\end{equation}
Here, we have assumed the anisotropy constant $\xi_{\rm b}$ to be unity, which corresponds to the isotropic case. The subscript `b' stands for burst. Keeping this equation in mind, we proceed towards deriving the Eddington luminosity from the equation (Pike et al., 2021)
\begin{equation}
L_{\rm Edd} = \frac{4\pi cGM}{\kappa_0}\bigg[1-\frac{2GM}{c^2 \rm R}\bigg]^{1/2} \times \bigg[ 1+ \bigg (\frac{kT}{39.2\: {\rm keV}}\bigg)^{0.86}\bigg] (1+X)^{-1}
\end{equation}
where M is the mass of the NS, G is the gravitational constant and \textit{c} is the speed of light. We have assumed a typical NS mass of 1.4 M$_\odot$. We  have considered the hydrogen mass fraction \textit{X}= 0 and the typical upper limit of the NS radius as R = 10 km. We use the value of $\kappa_0$ = 0.2 cm$^2$ g$^{-1}$, which is the opacity factor for pure He (Pike et al., 2021). Using the value of $kT_{\rm bb}$ for the T4 time bin of Table. 3, we find the value of Eddington luminosity $L_{\rm Edd}$ = 2.87 $\times$ 10$^{38}$ erg s$^{-1}$. Substituting this value of luminosity and choosing the value of $F_{\rm b}$ as the highest flux value from Table. 3, which corresponds to the T4 time bin in Eq. (\ref{first}), we achieve the distance $d$= 9.3 $\pm$ 0.4 kpc. For a range 0.5 $<$ $\xi_{\rm b}$ $<$ 2 (Kuulkers $\&$ van der Klis, 2000), we obtain $d$ in the range 6.6 - 13.2 kpc. We believe that to obtain source's physical parameters with greater accuracy, a refinement in the spectral analysis is required on a better data.

\section*{Acknowledgements}

This report work has utilized the softwares provided by the High Energy Astrophysics Science Archive Research Centre (HEASARC). The authors AN and BS acknowledge the financial support of ISRO under \textit{AstroSat} archival Data utilization program (No.DS\_2B-13013(2)/2/2019-Sec.2). This publication uses data from the \textit{AstroSat} mission of the Indian Space research Organization (ISRO), archived at the Indian Space Science Data Centre (ISSDC). AN acknowledges the Inter-University Center for Astronomy and Astrophysics (IUCAA), Pune for the periodic visits, the hospitality and the facilities provided to him to make possible the major part of the work. This work has been performed utilizing the calibration data-bases and auxillary analysis tools developed, maintained and distributed by AstroSat-SXT team with members from various institutions in India and abroad. This research has made use of MAXI data provided by RIKEN, JAXA and the MAXI team (Matsuoka et al., 2009). BS also acknowledges the visiting associateship program at IUCAA and is grateful to IUCAA for hospitality during his visits where part of this work was done. AN thanks Dr. M. Pahari, Royal Society SERB International fellow for his useful suggestions in this work. AN personally thanks the scholar mates from Jamia Millia Islamia University, Delhi and other fellow mates who helped him resolve issues faced in this work. BS would like to acknowledge discussions held with Dr. Anjali Rao during the initial stages of this work. Authors are also indebted to the reviewer for providing beneficial suggestions and constructive comments, which has greatly enhanced the quality of the work.

%%References section
\def\apj{ApJ}
\def\mnras{MNRAS}
\def\aap{A\&A}
\def\apjl{ApJL}
\def\baas{Bulletin of the AAS}
\def\physrep{PhR}
\def\aapr{A\&AR}
\def\prl{Ph.~Rev.~Lett.}
\def\apjs{ApJS}
\def\pasa{PASA}
\def\pasj{PASJ}
\def\japa{JApA}
\def\nat{Nature}
\def\memsai{MmSAI}
\def\aj{AJ}
\def\aaps{A\&AS}
\def\iaucirc{IAU~Circ.}
\def\sovast{Soviet~Ast.}
\def\apss{Ap\&SS}
\def\procspie{Proc.~SPIE}
\def\ssr{Space Sci. Rev}
\def\na{New Ast.}
\def\araa{ARAA}

\end{document}